\documentclass[english,aps,pre,nofootinbib,reprint,amssymb,showpacs,superscriptaddress,nobalancelastpage]{revtex4-1}
\usepackage[T1]{fontenc}
\usepackage[latin9]{inputenc}
\usepackage{color}
\usepackage{verbatim}
\usepackage{amsmath}
\usepackage{graphicx}
\usepackage{babel}
\begin{document}

\title{Parameters of state in the global thermodynamics of binary ideal gas mixtures in a stationary  heat flow}
\author{Anna Macio\l ek}
\email{equal contribution;amaciolek@ichf.edu.pl}
\affiliation{Institute of Physical Chemistry, Polish Academy of Sciences Kasprzaka
44/52, 01-224 Warszawa}
\affiliation{Max-Planck-Institut f{\"u}r Intelligente Systeme Stuttgart, Heisenbergstr.~3,
D-70569 Stuttgart, Germany}

\author{Robert Ho\l yst}
\email{equal contribution;rholyst@ichf.edu.pl}
\affiliation{Institute of Physical Chemistry, Polish Academy of Sciences Kasprzaka
44/52, 01-224 Warszawa}

\author{Karol Makuch}
\affiliation{Institute of Physical Chemistry, Polish Academy of Sciences Kasprzaka
44/52, 01-224 Warszawa}

\author{Konrad Gi\.zy\'nski}
\affiliation{Institute of Physical Chemistry, Polish Academy of Sciences Kasprzaka
44/52, 01-224 Warszawa}

\author{Pawe\l{} J. \.{Z}uk}

\affiliation{Institute of Physical Chemistry, Polish Academy of Sciences Kasprzaka
44/52, 01-224 Warszawa}
\affiliation{Department of Physics, Lancaster University, Lancaster LA1 4YB, United Kingdom}

\date{\today}
\begin{abstract}

We formulate the first
 law of global thermodynamics for stationary states of the binary
ideal gas mixture subjected to heat flow.
We map the non-uniform system onto the uniform one and show that 
the internal energy $U(S^*,V,N_1,N_2,f_1^*,f_2^*)$ is the function of the following parameters of state: 
a non-equilibrium entropy $S^*$, volume $V$, number of 
particles of the first component, $N_1$, number of particles of the second component $N_2$ and the
renormalized degrees of freedom. The parameters $f_1^*,f_2^*$, $N_1, N_2$ satisfy the relation  $x_1f_1^*/f_1+x_2f_2^*/f_2=1$ ($f_1$, where $x_i$ is the fraction of $i$ component, and $f_2$ are the degrees of freedom for each component respectively).  Thus only 
5 parameters of state describe the non-equilibrium state of the binary mixture in the heat flow. 
 We calculate the non-equilibrium entropy $S^{*}$ and new thermodynamic parameters of state $f_1^*, f_2^*$ explicitly. 
The latter are responsible for heat generation due to the concentration gradients. 
The theory reduces to equilibrium thermodynamics, when the heat flux goes to zero. 
As in equilibrium thermodynamics, the steady-state fundamental equation also leads to the thermodynamic Maxwell relations
for measurable steady-state properties.

\end{abstract}
\maketitle

\section{Introduction}

In the classical thermodynamics the internal energy, $U(S,V,N)$, of a one-component ideal gas  is a function of three parameters: entropy, $S$, volume $V$ and the number of particles $N$. Each parameter of state represents one independent way of system's energy exchange with the external world. 
For fixed $N$ there are two ways of energy change: heat and work. The infinitesimal change of the internal energy satisfies the equation: $dU=TdS-pdV$, where the first term is the heat and the last is the work term. Untill recently no such description was available for systems in non-equilibrium states, subjected to energy flow. 

The classical theory of irreversible (non-equilibrium) thermodynamics \cite{Groot_Mazur_Non-equilibrium_thermodynamics} is based on three differential non-linear equations representing conservation of mass, momentum (Navier-Stokes equation) and energy. These conservation laws are supplemented by the 
assumption of the local equilibrium,  corresponding local equations of state, and constitutive relations between fluxes and thermodynamic forces. The solutions of these equations
are given in terms of velocity, $v({\bf r},t)$, temperature, $T({\bf r},t)$ and number density of particles, $n({\bf r},t)$ profiles. In the stationary states the profiles and fluxes depend only on the position in space ${\bf r}$, but not on time, $t$.
The assumption of the local equilibrium is crucial in the formulation of the local equations of state. 
For the ideal gas it is justified, but contested for interacting systems \cite{Rubi}. The assumption of local equilibrium for the ideal gas
is valid for small  temperature gradient i.e. such that,
$l_{fp}\left|\nabla T\right|/T\ll1,$ for the mean free path of the
molecules, $l_{fp}$ \cite{Nonequilibrium_thermodynamics_and_its_statistical_foundations_H_J_Kreuzer}.
At the pressure of $1$ bar at room temperature, the mean free path
is of the order of $l_{fp}\approx 100$nm. In such conditions the assumption of local equilibrium breaks down 
only for temperature gradients  higher than $10^{7}K/cm$. Thus this assumption is even satisfied inside the Sun! 
This local description contains the first law of thermodynamics, but in a different form than that given in the equilibrium thermodynamics and based on a few \textit{global} parameters of state.
In our recent paper \cite{JCP22} we provided the latter, i.e.,   global thermodynamic  description for the ideal gas in a heat flow.

We have previously \cite{JCP22} studied the one-component  ideal gas in a heat flow between two parallel walls at distance $L$,  kept at two different temperatures $T_1(z=0)>T_2(z=L)$. The local equilibrium gives the local pressure, $p(z)$,  and the internal energy per unit volume $u(z)$:
\begin{equation}
p\left(z\right)=k_{B}T\left(z\right)n\left(z\right),\label{eq:eq of state pressure}
\end{equation}
and
\begin{equation}
u\left(z\right)=\frac{3}{2}n\left(z\right)k_{B}T\left(z\right),\label{eq:energy eos}
\end{equation}
with Boltzmann constant $k_{B}$, particle
number density $n\left(z\right)$, and the temperature $T\left(z\right)$
at position $z$.  
We have shown rigorously that when we integrate both equations over the volume of the system we can formulate the global thermodynamics with the internal energy as a function of a few parameters of state. After integration we get:
\begin{equation}
pV=Nk_{B}\frac{T_{2}-T_{1}}{\log\frac{T_{2}}{T_{1}}}.\label{eq:p by T1 T2}
\end{equation}
and 
\begin{equation}
U=\frac{3}{2V}Nk_{B}\frac{T_{2}-T_{1}}{\log\frac{T_{2}}{T_{1}}}.\label{eq:u by T1 T2}
\end{equation}
After definition of system's temperature, $T^*$,
\begin{equation}
T^*=\frac{T_{2}-T_{1}}{\log\frac{T_{2}}{T_{1}}}.\label{eq:T by T1 T2}
\end{equation}
In this way we made a mapping of the non-uniform system into the uniform one.
It is worth mentioning that the local equilibrium
assumption is not needed. If we had different local equations of state we could still integrate them over the volume and perform aforementioned mapping. 
We observe that obtained equations have the same form as in equilibrium  for the ideal gas at temperature $T^*$. 
We demanded such form because after the mapping we treat the system as the uniform one as in equilibrium. Moreover the obtained equations of state must reduce the equilibrium equations of state when the heat flux is zero. Now we define the internal energy as a function of three parameters of state $U(S^*,V,N)$, with the thermodynamic relation:
\begin{align}
\left(\frac{\partial S^{*}}{\partial U}\right)_{V,N} & =\frac{1}{T^{*}},\label{eq:temp st by partial}\\
\left(\frac{\partial S^{*}}{\partial V}\right)_{U,N} & =\frac{p}{T^{*}}.\nonumber 
\end{align}
This mapping gives us the same formal structure as we know from equilibrium. The entropy $S^*$ is responsible for the {\bf net heat} that enters or
leaves the system \cite{oono1998steady,nakagawa2019global,chiba2016numerical} and changes the internal energy. In general the heat flows through the system all the time without changing the internal energy. Upon any process we would like to know how much of the heat transferred to the system changes the internal energy. This heat is called net heat and is given in the differential form by  $T^*dS^*$.

$S^*$ is only part of the total entropy of the system. The total entropy is given by:
\begin{align}
S_{\text{tot}}\left(U,V,\frac{T_{2}}{T_{1}}\right) & =S^{*}\left(U,V\right)+\Delta S\left(U,V,\frac{T_{2}}{T_{1}}\right),\label{eq:s star and total}\\
\Delta S\left(U,V,T_{2}/T_{1}\right) & =Nk_{B}\log\left[\left(\frac{T_{2}}{T_{1}}\right)^{5/4}\left(\frac{\log\frac{T_{2}}{T_{1}}}{\frac{T_{2}}{T_{1}}-1}\right)^{5/2}\right].\nonumber 
\end{align}
Only $S^{*}$ governs net
heat in the system (heat absorbed/released in the system).
$\Delta S$, controls the dissipative background and solely depends
on the entropy production given by \cite{Groot_Mazur_Non-equilibrium_thermodynamics}
$\sigma=-A\int_{0}^{L}dz\,\kappa\nabla T\left(z\right)\cdot\nabla\frac{1}{T\left(z\right)}=\frac{A\kappa}{L}\left(\frac{T_{2}}{T_{1}}+\frac{T_{1}}{T_{2}}-2\right)$, where $\kappa$ is the heat conductivity.
The difference between the total entropy and $S^{*}$ vanishes, $\Delta S\left(U,A,L,T_{2}/T_{1}\right)\to0$,
when the system approaches the equilibrium state, $T_{2}/T_{1}\to1$.
Therefore, $S^{*}$ becomes in this limit the equilibrium entropy. Nonetheless the formal dependence of $S^*$  
on $U$ and $V$ at non-equilibrium state in a heat flow is the same as at equilibrium.

In this paper we want to apply the same mapping procedure to the binary mixture of  ideal gases. The big difference between previous work and the current one is the fact that apart from the number density profile and the temperature profile we have additional profiles of the number densities of each component. 
As we shall see, these additional profiles lead to new parameters of state in the non-equilibrium state, which have no direct counterpart at equilibrium.
The purpose of this work is to formulate the first law of global thermodynamics for ideal gas binary mixture in the heat flow.
The paper is organized as following: in section II we recall the equilibrium properties of the ideal gas mixture. 
In section III we discuss this mixture enclosed between two parallel walls at different temperatures and  
 solve the equations of irreversible thermodynamics.   In section IV we define all parameters of state for the mixture and perform  the mapping of non-uniform system into the uniform one. 
 We introduced the first law of non-equilibrium thermodynamics, which follows from these parameters, and discuss some 
 of its consequences in section V.
 We determine the  discuss the results in section VI.

\section{Preliminaries}

We consider a binary mixture of ideal gases enclosed between two parallel walls  separated from each other by a distance
 $L$ in the $z$ direction.
The volume of a system is $V=A L$, where  $A$ is an area in the $x-y$ plane.
The components of the mixture have  number densities $n_1$ and $n_2$, such that  $n=N/V= n_1+n_2$ is the total density.
The equation of state of  ideal gas  at pressure $p$ and  temperature $T$ 
$p= nk_{B}T$, where $k_{B}$ is  the Boltzmann constant, can be written as  a sum of partial pressures $p_i$ 
\begin{equation}
\begin{aligned} 
p&=&p_1+p_2 = n_1 k_BT + n_2 k_BT \nonumber \\
&=&x_1 n k_BT + (1-x_1)n k_BT,\label{eq:Dalton}
\end{aligned}
\end{equation}
which is  the Dalton's law~\cite{Callen}.   $x_i=n_i/n$ is the fraction of $i$ component;   $x_1+x_2=1$ in the absence of chemical reactions.

The  internal energy  density (per volume)  is  the sum of the internal energy density $u_i$ ($i=1,2$) of the two components separately considered:
\begin{equation}
u = u_1+u_2,  \label{eq:energy tot}
\end{equation}
From  the classical equipartition theorem applied to ideal gas it follows that
\begin{equation}
u_i = \frac{f_i}{2} n_i k_BT = \frac{f_i}{2} p_i, \quad \quad i=1,2\label{eq:energy}
\end{equation}
where $f_i, i=1,2$   are translational and rotational degrees  of freedom of the $i$ component.
 $f_i$ takes the value  3 for monoatomic,  5  for diatomic, and  6  for polyatomic gas.
Often, the ideal gas law (\ref{eq:energy}) is  written using dimensionless specific heat capacity at constant volume $c^{(i)}_v=\frac{f_i}{2}$.

The total energy and entropy are given by 
\begin{equation}
 \label{eq:Int}
 U=\int_V u d^3r \quad \mathrm{and} \quad S=\int_V ns d^3r,
\end{equation}
(which reduces to $U=Vu$ and $S=Ns$ for homogeneous systems) 
where the equilibrium entropy  density $ns= ns^{(0)} + n s_{mix}$  of a binary mixture is 
\begin{equation}
 \label{eq:entropy}
 n\cdot s = ns^{(0)} + k_B\left[ n \ln n -n_1 \ln n_1 - n_2 \ln n_2\right].
 \end{equation}
Here $ ns^{(0)} = n_1 s_1^{(0)} + n_2 s_2^{(0)}$, where 
 $s_i^{(0)} \quad (i=1,2)$ is the entropy per  particle of the two components separately considered. $s_{mix}$ is
the mean-field expression for the mixing entropy.
For ideal gas one has
\begin{equation}
 \label{eq:s_idealgas}
 \frac{s_i^{(0)}}{k_B}  = \frac{f_i}{2} +1 +\frac{f_i}{2} \ln \left(\frac{2\Phi_iu_i}{f_ik_Bn_i^{1+2/f_i}}\right),
\end{equation}
where $\Phi_i$  is independent of the thermodynamic state of the gas 
and has dimension of $n^{2/f_i}/T$. 
For a monoatomic ideal gas, a quantum mechanical theory of Sacur-Tetrode predicts that the constant $\Phi_i$ 
depends only upon the mass of the gas particle~\cite{Sackur,Tetrode}.

From fundamental thermodynamic relation in terms of entropy, the differential $dS$ has the following form
\begin{equation}
 \label{eq:diff_ent}
 dS=\frac{1}{T}dU+\frac{p}{T} dV-\frac{\mu_1}{T}dN_1-\frac{\mu_2}{T}dN_2.
\end{equation}
Using expressions for the entropy (Eqs.~(\ref{eq:entropy}) and (\ref{eq:s_idealgas})), the chemical potential 
of each component of a mixture is obtained 
as $\frac{\mu_i}{T} = -\left(\frac{\partial S}{\partial N_i}\right)_{U,V,N_{j\ne i}}=-\left(\frac{\partial ns}{\partial n_i}\right)_{u,n_{j\ne i}}$, which gives
\begin{equation}
 \label{eq:chem_pot}
 \frac{\mu_i}{T} = \frac{\mu_i^{(0)}}{T} + k_B\ln x_i,
\end{equation}
where $\frac{\mu_i^{(0)}}{T} = -  \left(\frac{\partial n_i s_i^{(0)}}{\partial n_i}\right)_{u_i,n_{j\ne i}}$ is  the chemical potential of the component $i$ individually considered;
it depends only on temperature $T$ and pressure $p$:
\begin{equation}
 \label{eq:che_pot_0}
 \frac{\mu_i^{(0)}}{T}(T,p)=-k_B\frac{f_i}{2}\ln \left(\frac{\Phi_ik_B^{2/f_i}T^{1+2/f_i}}{p^{2/f_i}}\right )
\end{equation}
We define the chemical potential difference  as $\frac{\mu}{T} =- \left(\frac{\partial S}{\partial \varphi}\right)_{u,n}$, where $\varphi = n_1-n_2$. 
This gives
\begin{equation}
 \label{eq:chem_diff}
 \frac{\mu}{T} = \frac{\mu_1-\mu_2}{2T} =  \frac{\mu_1^{(0)}-\mu_2^{(0)}}{2T}+\frac{k_B}{2} \ln\frac{x_1}{1-x_1}.
\end{equation}

In the absence of external fields, the equilibrium ideal gas is homogeneous.

\section{Ideal gas mixture in heat flow}

\begin{figure}[bp]
\includegraphics[width=8.5cm]{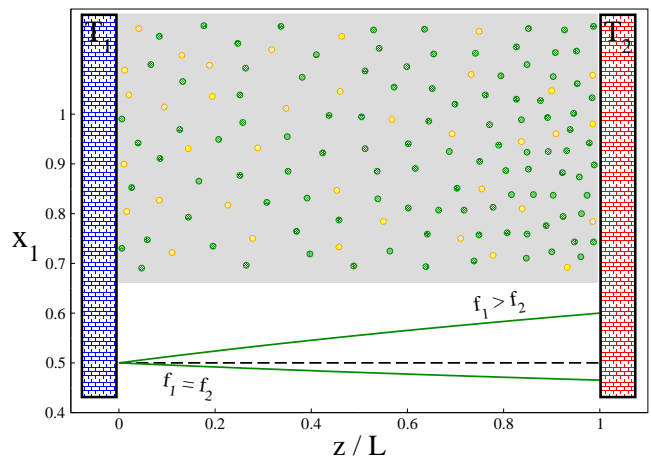}
\caption{Schematic illustration of a binary  ideal gas mixture between two parallel walls held at different temperatures $T_2$ and $T_1$. 
The resulting temperature gradient induces a gradient of  fractions $x_i$ of the mixture components. 
Solid (green) lines indicate profiles of  $x_1(z/L)$. Two cases are shown: (i) $f_1>f_2$ for which the profile is
given by Eq.~(\ref{eq:xT}) and (2) $f_1=f_2$ for which the profile is given by  Eq.~(\ref{eq:xT0_Soret_sol}).
For the latter case, we took  $\alpha_{12}  = 0.3432$, which corresponds to the Ne-He pair at $T_1=300$ K. 
In both cases the reduced temperature gradient $s=(T_2-T_1)/T_1$ was chosen to be 0.5 and $N_1=N_2$.
}
\label{fig:1}
\end{figure}

Now, we introduce heat flow into the system by setting different temperatures on the walls, i.e.,  
\begin{align}
T\left(z=0\right) & =T_{1},\nonumber \\
T\left(z=L\right) & =T_{2}.\label{eq:temp bc external walls}
\end{align}
In the non-equilibrium state induced by this boundary condition, the system becomes inhomogeneous and one has to consider spatially varying temperature $T\left(\mathbf{r}\right)$,
pressure $p\left(\mathbf{r}\right)$, density $n\left(\mathbf{r}\right)$ and difference in number densities $\varphi\left(\mathbf{r}\right)$.
 In the hydrodynamic limit, the time evolution of the binary mixture  is given by the conservation laws for $n$, $\varphi$,  momentum, and energy supplemented by  the assumption of local equilibrium, relations between
thermodynamic forces and fluxes,  and thermodynamic equations of state~\cite{Groot_Mazur_Non-equilibrium_thermodynamics}. 
Note that  ideal gas  satisfies  the local equilibrium  exactly. This fact describes the strict absence of spatial correlations between particles, which is, of course, not the case for  non-ideal systems.

We focus here on a stationary state with vanishing gas velocity field and  constant pressure  across the system. The latter, 
together with $n\left(\mathbf{r}\right)$ and $\varphi\left(\mathbf{r}\right)$, follows from conservation of mass and momentum.
The assumptions $\mathbf{v}=0$ and $p\left(\mathbf{r}\right)=p$ are  in agreement with  our previous  simulation studies ~\cite{babin2005,holyst2008,holyst2019} 
(for  gas-liquid evaporating systems~\cite{babin2005,holyst2008} and for Lennard-Jones fluid  volumetrically heated~\cite{holyst2019}), which show that mechanical equilibrium is established very fast (in comparison to heat flow).
Thus, the  balance equations for $\varphi$ and energy simplify to
\begin{equation}
\begin{aligned} \label{eq:dyn_con_en}
  \frac{\partial \varphi}{\partial t} & = &- 2\partial_i J_i^d, \\
 \frac{\partial e}{\partial t} & = & - \partial_iJ_i^Q,
 \end{aligned}
   \end{equation}
where $\mathbf{J}^{d}\equiv \mathbf{J}^{d,1}=-\mathbf{J}^{d,2}$ is the diffusion current and $\mathbf{J}^{q}$ is the heat current. 
We choose the following  phenomenological expressions for $\mathbf{J}^{d}$ and $\mathbf{J}^{q}$ ~\cite{Groot_Mazur_Non-equilibrium_thermodynamics}:
\begin{equation}
\begin{aligned} \label{eq:current}
 \mathbf{J}^d &=& -\mathcal{L}_{11} \mathbf{\nabla}\left(\frac{\mu}{T}\right) + \mathcal{L}_{12} \mathbf{\nabla}\left(\frac{1}{T}\right), \\
\mathbf{J}^q &=& -\mathcal{L}_{21} \mathbf{\nabla}\left(\frac{\mu}{T}\right) + \mathcal{L}_{22} \mathbf{\nabla}\left(\frac{1}{T}\right)
\end{aligned}
\end{equation}
In steady state, $ \mathbf{J}^d =0$  because the system has no contact with particles reservoir. 

First, we 
ignore the thermodiffusion effect, i.e., we assume that $\mathcal{L}_{12}=0$. We also neglect the other cross-term by setting  $\mathcal{L}_{21}=0$.
As a result,  Eq.~(\ref{eq:current}) reduces to 
\begin{equation} \label{eq:c1}
\mathbf{\nabla}\left(\frac{\mu}{T}\right) =0
\end{equation}
and
\begin{equation}
\mathbf{\nabla} T = const \Rightarrow \mathbf{\nabla}^2 T=0.
\end{equation}
\vspace{0.3cm}
The latter yields a linear temperature profile (the system is translationally invariant in $x,y$ directions)
\begin{equation} \label{eq:prof}
 T(z) =T_1+(T_2-T_1) \frac{z}{L}.
\end{equation}
From Eq.~(\ref{eq:c1}) we determine the relationship  between the local fraction $x_1(z)$  of the component 1 and the temperature.
The local equilibrium allows us to use the local form of  Eq.~(\ref{eq:chem_diff}) for $\mu/T$; it depends on  $T(z), x_1(z), x_2(z)$,
and the pressure $p$, which is constant throughout the system.
Thus
\begin{equation}
 \label{eq:chem_der}
 \mathbf{\nabla}\left(\frac{\mu}{T}\right) = \frac{h_1-h_2}{2}\mathbf{\nabla}\frac{1}{T}+\frac{\mathbf{\nabla}_T\left(\mu_1-\mu_2\right)}{2T},
\end{equation}
where $h_1$ is the enthalpy per particle and $\mathbf{\nabla}_T$ is the gradient at constant temperature;
it involves the derivative of the chemical potential with respect to composition, i.e., 
\begin{equation}
 \label{eq:chem_der_x}
 \mathbf{\nabla}_T\left(\mu_1-\mu_2\right)= \frac{\partial\mu_1}{\partial x_1}\mathbf{\nabla}x_1-\frac{\partial\mu_2}{\partial x_2}\mathbf{\nabla}x_2.
\end{equation}
Assuming  that the reference state has temperature $T_1$ and using the fact that $h_i=H_i/N_i = (f_i/2 +1)k_BT$ for the ideal gas we  find 
\begin{equation}
 \label{eq:xT0}
 \nabla\ln \frac{x_1(z)}{x_2(z)} = \nabla \ln \tilde T(z)^{\frac{f_1-f_2}{2}},
\end{equation}
where $\tilde T(z) = T(z)/T_1$. 
With  $x_1(z)=1-x_2(z)$ this  gives
\begin{equation} \label{eq:xT}
x_1(z) = \frac{N_1\tilde T(z)^{(f_1-f_2)/2}}{N_2+N_1\tilde T(z)^{(f_1-f_2)/2}}.
\end{equation}
Note that if $f_1=f_2$, temperature inhomogeneity does not induce spatial variation of the composition, unless we take into account the effect of thermodiffusion. In that case we have
generally
\begin{equation}
 \label{eq:xT0_Soret}
 \nabla \ln\frac{x_1(z)}{x_2(z)} =  \left[\frac{f_1-f_2}{2}-\frac{2\mathcal{L}_{12}}{k_B\mathcal{L}_{11}T(z)}\right]\nabla \ln \tilde T(z).
\end{equation}
Many numerical results as well as measurements of  thermodiffusion coefficients are reported in the literature (see e.g.~\cite{transport,Chapman,consolidated} and references therein).
They indicate that these coefficients  are  functions of composition and temperature.
Within the framework of the generalized Stefan-Maxwell thermodiffusion equations
 for $f_1=f_2$,  one obtains exactly the same equation as above,  but with  $2\mathcal{L}_{12}/(\mathcal{L}_{11}k_BT)$  replaced by $\alpha_{12}=\mathcal{A}_{12}/\mathcal{D}_{12}$. 
 $\mathcal{A}_{ij}$ is  the Newman-Soret thermal diffusion coefficient and $\mathcal{D}_{ij}$ 
is the Stefan-Maxwell diffusivity (see e.g., Eq.~(6.8) from  Ref.~\cite{consolidated}). 
It is argued that coefficients  $\mathcal{A}_{ij} = D^T_i/\rho_i-D^T_{j}/\rho_j$, where $D^T_{i}$ is the thermal diffusivity coefficient  of component $i$ and $\rho_i$ is  its mass density,  
are more approximately constant with respect to composition in binary systems~\cite{Newman}.
Assuming that   the ratio $\alpha_{12}$  does not depend on both temperature and composition,
the solution of Eq.~(\ref{eq:xT0_Soret}) for the profile $x_1(z)$ is 
\begin{equation}
 \label{eq:xT0_Soret_sol}
 x_1(z)=\frac{N_1\tilde T(z)^{(f_1-f_2)/2-\alpha_{12}}}{N_2+N_1\tilde T(z)^{(f_1-f_2)/2-\alpha_{12}}}.
\end{equation}
This solution clearly shows  that the  difference $f_1-f_2$ is  a relevant parameter.
An analogy can be drawn here with the order parameter: for  non-zero  $f_1-f_2$, 
the profile $x_1(z)$ is \textit{qualitatively} different from that for the zero difference $f_1-f_2$. Specifically, for $f_1-f_2 > 0$, the concentration of the first component
of the mixture (with more degrees of freedom)  increases towards the hotter wall, while 
for  $f_1=f_2$ the opposite is true (see Fig.~\ref{fig:1}). This also  holds  if, due to the concentration and temperature dependence of the thermodiffusion coefficients,  the functional form of $x_1$-dependence 
on  $T$ is different than that given by Eq.~(\ref{eq:xT0_Soret_sol}).
For noble gas pairs at $T=300$ K, the value of $\alpha_{12}$ is equal to
 0.3432 for Ne-He pair, 0.1741 for Ar-Ne pair, or  0.0262 for Xe-Kr pair \cite{Marrero,Taylor}.
 If  we assume that Onsager coefficients $\mathcal{L}_{12}$ and $\mathcal{L}_{11}$ are independent on the concentration and temperature~\cite{onsager1931reciprocal},
the solution of Eq.~(\ref{eq:xT0_Soret}) becomes
\begin{equation}
 \label{eq:xT0_Soret_sol_1}
 x_1(z)=\frac{N_1\tilde T(z)^{(f_1-f_2)/2}}{N_2e^{\frac{2\mathcal{L}_{12}}{k_BT_1\mathcal{L}_{11}}\left(1-\frac{1}{\tilde T(z)}\right)}+N_1\tilde T(z)^{(f_2-f_2)/2}}.
\end{equation}

The density profile follows from the local equation of state and the condition of constant pressure at steady state,
$n\left(z\right)=p/k_{B}T\left(z\right)$. For a given number 
of particles, $N=A\int_{0}^{L}dz\,n\left(z\right)$,
this  determines pressure  as
\begin{equation}
p=\frac{N}{V}k_{B}\frac{T_{2}-T_{1}}{\log\frac{T_{2}}{T_{1}}}.\label{eq:pT1T2}
\end{equation}
Finally, the total energy of a mixture  is the sum of the energies of the both components of the mixture (see Eqs.~(\ref{eq:Int}), (\ref{eq:energy tot}), and (\ref{eq:energy}))
\begin{equation}\label{eq:toten}
\begin{aligned}
 U  = U_1+U_2 &=& A\frac{f_1}{2}k_B\int_0^L n(z) x_1(z) T(z)dz  \\ 
 &+& A\frac{f_2}{2}k_B\int_0^L n(z) x_2(z) T(z)dz.
 \end{aligned}
\end{equation} 
Thus, the set of independent  parameters controlling  the stationary state is $T_{1},T_{2},A,L,N,N_1,f_1,f_2$.

\section{Non-equilibrium parameters and functions of state}

%For the system described in the preceding section we construct a nonequilibrium state function 
One can  rewrite the total energy (per volume)  as follows
\begin{equation}
\begin{aligned}\label{eq:totenstar}
 u = \frac{U}{V} = \frac{f_1^*}{2}n_1k_B T^* + \frac{f_2^*}{2}n_2k_B T^*,
 \end{aligned}
\end{equation} 
where $n_i=N_i/V=(1/L)\int_0^L n_i(z)dz$. This has the same form as in equilibrium but with temperature $T$ replaced by $T^*$ and  $f_1, f_2$
replaced by  parameters $f_1^*,f_2^* $, which we call the ``effective'' degrees of freedom. From Eq.~(\ref{eq:toten}) it follows that 
\begin{equation} \label{eq:f1}
 f_1^*=f_1\frac{\frac{1}{L}\int_0^L n(z) x_1(z) \frac{k_BT(z)}{2}dz}{n_1\frac{k_BT^*}{2}}= f_1 p \frac{\frac{1}{L}\int_0^L x_1(z)dz}{n_1k_BT^*} 
\end{equation} 
and 
\begin{equation} \label{eq:f2}
 f_2^*=f_2\frac{\frac{1}{L}\int_0^L n(z)x_2(z) \frac{k_BT(z)}{2}dz}{n_2 \frac{k_BT^*}{2}}= f_2 p \frac{\frac{1}{L}\int_0^Lx_2(z)dz}{n_2k_BT^*} 
\end{equation} 
where  we have used  $n(z)T(z)= p/k_B = const$. 
The parameter $T^*$ can be determined from the requirement that the pressure 
\begin{equation}\label{eq:pressure1}
 p=k_B\frac{1}{L}\int_0^L \Big(n_1(z)T(z) +n_2(z)T(z)\Big)dz = p_1 + p_2.
\end{equation} 
 Then, Eq.~(\ref{eq:pressure1}) can be written as
\begin{equation} \label{eq:pressure2}
 p=n_1k_BT^* + n_2k_BT^*,
\end{equation} 
which is to say that partial pressures in steady state are related to the non-equilibrium energies in the same way as in equilibrium (see Eq.~(\ref{eq:energy}))
but with  $f_i$ replaced by $f_i^*$. Solving Eq.~(\ref{eq:pressure2}) with the use of Eq.~(\ref{eq:pT1T2}) we find
\begin{equation} \label{eq:Tstar}
 T^*= \frac{p}{k_B}\frac{1}{n_1+n_2}= \frac{p}{k_B}\frac{V}{N}=  \frac{T_2-T_1}{\ln \frac{T_2}{T_1}}
\end{equation}
$T^*$ can be interpreted as an ``average'' temperature in the non-equilibrium steady state.
Eliminating $p$ in expressions for  $f_1^*, f_2^*$ we find:
\begin{equation}
\begin{aligned}
\label{eq:f1af2}
 & f_1^*= \frac{f_1}{x_1}\frac{1}{L}\int_0^L x_1(z)dz, \\
 &f_2^*= \frac{f_2 }{x_2}\frac{1}{L}\int_0^L x_2(z)dz.
 \end{aligned}
\end{equation} 
Using $x_2(z)=1-x_1(z)$ we can see that the new parameters of state  $f_1^*, f_2^*$  are not independent but obey the following relation:
\begin{equation}
 \label{eq:f1vf2}
 \frac{f_1^*}{f_1}x_1+\frac{f_2^*}{f_2}x_2=1.
\end{equation}
We note that even if both components of the mixture have the same number of degrees of freedom $f_1=f_2$, the effective parameters $f_1^*, $ are not equal in non-equilibrium steady states.
Using Eqs.~(\ref{eq:xT}) and (\ref{eq:prof}) for the profiles of the  fraction $x_1(z)$ of the component 1  and temperature $T(z)$, 
we can express the steady state variable $f_1^*$  in terms of the control parameters.
The explicit formulas can be obtained for the case of $f_1\ne f_2$ in terms of the special functions as  presented in Appendix~\ref{app:A}. 
For small, reduced temperature gradients
$s = (T_2-T_1)/T_1$ and for $N_1=N_2$ we find
\begin{equation}
\label{eq:f1a}
f_1^*\approx \frac{f_1}{x_1}\left(\frac{1}{2} + \frac{f_1-f_2}{16}s\right) +O(s^2).
\end{equation}
If  $N_1 \ne N_2$, the expression for $f_1^*$ in this limit  is more complicated, but has a similar structure
in terms of  dependence on $s$. The coefficients  are functions of $N_1/N_2$ and $f_1, f_2$
but not of $T_1$ as shown in Appendix~\ref{app:A}. 
In the case of $f_1 = f_2$, the integral over the fraction  profile $x_1(z)$ given by 
Eq.~(\ref{eq:xT0_Soret_sol}) has the same form as for $f_1\ne f_2$ (given in Appendix~\ref{app:A}),
but with $f_1-f_2$ replaced by $-\alpha_{12}$. For a profile given by Eq.~(\ref{eq:xT0_Soret_sol_1})
the integral cannot be expressed analytically.

In the next step in the construction of  global thermodynamics, we note that because the equations (\ref{eq:totenstar}) and (\ref{eq:pressure2})   have the same structure as the equilibrium equation of state, we may formally write:
\begin{eqnarray}
\label{eq:non_entropy}
S^* = (S^*_1)^{(0)}+(S^*_2)^{(0)} + S^*_{mix},
\end{eqnarray}
with
\begin{eqnarray}
\label{eq:non_entropy1}
&&\frac{(S^*_i)^{(0)}}{N_ik_B} = \frac{f_i^*}{2}+1 \nonumber \\
&+&\frac{f_i^*}{2 } \ln \left[\frac{2\Phi_iU}{k_B(N_1f_1^*+N_2f_2^*)}\left(\frac{V}{N_i}\right)^{2/f_i^*}\right],
\end{eqnarray}
and 
\begin{eqnarray}
\label{eq:non_entropy2}
\frac{S^*_{mix}}{k_B} =-N_1\ln \frac{N_1}{N}-N_2\ln \frac{N_2}{N}.
\end{eqnarray}
Equation (\ref{eq:non_entropy}) has the functional form of  the equilibrium fundamental relation for a binary ideal gas mixture  with $f_i$ replaced by $f_i^*$ 
(compare Eqs~(\ref{eq:entropy}) and (\ref{eq:s_idealgas})).
We will treat Eqs~(\ref{eq:non_entropy})-(\ref{eq:non_entropy2}) as a definition of the \textit{non-equilibrium} steady state entropy $ S^*(U,V,N_1,N_2,f_1^*,f_2^*)$
that also provides the \textit{fundamental relation for the non-equilibrium steady state}.
Note that in comparison with the equilibrium entropy, $S^*$ depends on additional two state parameters $f_1^*$ and $f_2^*$. 
As a consequence, for the difference $dS^*$ we find:
\begin{eqnarray}
\label{eq:entr_diff}
dS^*&=&\frac{dU}{T^*}+\frac{p}{T^*}dV \nonumber \\
&-&\frac{\mu_1^*}{T^*}dN_1-\frac{\mu_2^*}{T^*}dN_2-\frac{A_1}{T^*}df_1^*-\frac{A_2}{T^*}df_2^*,
\end{eqnarray}
where
\begin{equation}
 \label{eq:mu*}
 \begin{aligned}
 &\frac{\mu_i^*}{k_BT^*} = -\left(\frac{\partial S^*}{\partial N_i}\right)_{U,V,N_{j\ne i},f_1^*,f_2^*} \\
 &=-\frac{f_i^*}{2 } \ln \frac{\Phi_ik_BT^*}{n_i^{2/f_1^*}} +\ln x_i
 \end{aligned}
\end{equation}
and 
\begin{equation}
 \label{eq:A_i}
 \frac{A_i}{T^*}=-\left(\frac{\partial S^*}{\partial f_i^*}\right)_{U,V,N_1,N_2,f_{j\ne i}^*}=-\frac{N_i}{f_i^*}\left(\ln x_i - \frac{\mu_i^*}{T^*}\right).
\end{equation}

Equations (\ref{eq:f1a}) and  (\ref{eq:Tstar}) provide the effective parameters of state, $T^*$ and $f_1^*$ 
 for given control parameters $T_1, T_2$ and $N_1, N_2$ at fixed $N_1+N_2$.
\section{First law and its consequences}

We now consider the change of the total internal energy $dU$ in our system. From Eq.~(\ref{eq:entr_diff}) we have
\begin{eqnarray}
\label{eq:en_diff}
dU=&T^*dS^*-pdV +\mu_1^*dN_1+\mu_2^*dN_2 \nonumber \\
&+A_1df_1^*+A_2df_2^*.
\end{eqnarray}
As argued in Ref.~\cite{JCP22},  the change in internal energy $dU$  in the very slow process of moving from one steady state to another by varying the control  parameters (so  that the  pressure remains homogeneous) is given by
\begin{equation}
 \label{eq:en_balance}
 dU=\mkern3mu\mathchar'26\mkern-12mu dQ+\mkern3mu\mathchar'26\mkern-12mu dW,
\end{equation}
where $dU=\mkern3mu\mathchar'26\mkern-12mu dQ$ is the net heat entering the system during this transition, and for the \textit{ fixed} number of mixture components  the mechanical  work done is
\begin{equation}
 \label{eq:work}
\mkern3mu\mathchar'26\mkern-12mu dW =-pdV.
\end{equation}
The equation (\ref{eq:en_balance}) is therefore the first law of  thermodynamics for non-equilibrium steady states where
 the net heat identified by Eqs~(\ref{eq:en_balance}) and (\ref{eq:entr_diff}) (at constant $N_1$ and $N_2$) is
\begin{equation}
 \label{eq:netheat}
 \mkern3mu\mathchar'26\mkern-12mu dQ=dU+pdV=T^*dS^*+A_1df_1^*+A_2df_2^*.
\end{equation}
We note that the net heat flow during the transition between two steady states is a combination of  two exact differentials: the effective entropy $dS^*$ and the effective degrees of freedom  $df_1^*$
(from   Eq.~(\ref{eq:f1vf2}) it follows that  $df_2^* = -\frac{x_2f_1}{x_1f_2}df_1^*$). This is contrary to the equilibrium thermodynamics, in which  heat depends only on temperature and the change in entropy.

At equilibrium, the most experimentally available thermodynamic quantities are response functions such as  heat capacities, compressibility, susceptibility or chemical response functions.
They follow from the first law of thermodynamics and the equilibrium fundamental relationship. Using the symmetry of the second derivatives, i.e., Maxwell relations, one can express one response function
  in terms of others. This is very useful as, for example, the heat capacity can be determined by measurements of other quantities, such as isothermal compressibility~\cite{Callen}.

 Once we have established the fundamental relationship for steady states of binary mixtures of ideal gases, we can generalize the equilibrium response functions to steady-state response functions.
 First, we consider thermal response function, i.e.,  the heat capacity $C$, which  is a measure of amount of heat needed to raise the temperature of a system by a given amount.
 Generally, it is defined as derivative, $C=\mkern3mu\mathchar'26\mkern-12mu dQ/dT$, but depending on which independent variables are fixed during the measurements, one has  different heat capacities.
 For example,  the heat capacity at  constant volume and  number of components $\{N_j\}$  is defined as $C_V= \mkern3mu\mathchar'26\mkern-12mu dQ/dT_{\mid_{V,\{N_j\}}}$, whereas 
 $C_p= \mkern3mu\mathchar'26\mkern-12mu dQ/dT_{\mid_{p,\{N_j\}}}$ is the heat capacity at constant  pressure and $\{N_j\}$.
 While in the equilibrium state at a constant volume and  fixed  $\{N_j\}$ and $\{f_j\}$ we can change only the temperature of the system, 
 in the non-equilibrium steady state we have more possibilities:  we can independently vary $T_1$ and $T_2$, or equivalently $T_2$ and the reduced difference $s=(T_2-T_1)/T_1$. 
If one changes $T_2$ at fixed $s$ then the state parameter $f_1^*$  remains constant (see Appendix~\ref{app:A} and Eq.~(\ref{eq:f1a})) we can define steady-state heat capacities for constant volume and pressure as follows
\begin{equation}
\label{eq:hcV}
C^*_V=\frac{\mkern3mu\mathchar'26\mkern-12mu dQ}{dT^*}_{\mid_{V,\{N_j\}},f_1^*} 
\end{equation}
and
 \begin{equation}
\label{eq:hcp}
C^*_p=\frac{\mkern3mu\mathchar'26\mkern-12mu dQ}{dT^*}_{\mid_{p,\{N_j\}},f_1^*}. 
\end{equation}
Concerning mechanical response functions, we can generalize isothermal compressibility and thermal expansivity to the steady-states as follows
 \begin{equation}
\label{eq:iscom}
\kappa^*_{T,{N_j}}= -\frac{1}{V}\left(\frac{\partial V}{\partial p}\right)_{\mid_{T^*,\{N_j\}},f_1^*} 
\end{equation}
and
 \begin{equation}
\label{eq:thexp}
\alpha^*_{p,{N_j}}=\frac{1}{V}\left(\frac{\partial V}{\partial T^*}\right)_{\mid_{p,\{N_j\}},f_1^*}. 
\end{equation}
Because the fundamental equation of steady-state has the same form as in equilibrium, the Maxwell relations for the path with  $s=const$ and $f_1^*=const$ provide  connection
between the thermal and mechanical response function, which is the same as in the equilibrium, i.e.,
\begin{equation}
 \label{eq:rel}
 \kappa^*_{T,{N_j}}\left( C^*_p - C^*_V\right) = T^*V\left(\alpha^*_{p,{N_j}}\right)^2.
 \end{equation}
In order to determine  $C^*_p$ one need to measure the excess heat due to the small change of $T_2$,  which is in principle possible due to the recent  development of
the appropriate experimental techniques~\cite{Yamamoto}. The coefficient $\kappa^*_{T,{N_j}}$ should be measured by changing the pressure at fixed both temperatures $T_1$ and $T_2$,
while the coefficient $\alpha^*_{p,{N_j}}$ should be determined by varying $T_2$ at fixed reduced gradient $s$.

\section{Summary}
\label{sec:Sum}

The internal energy of the binary ideal gas mixture in a heat flow is the function of 5 parameters of state $U(S^*,V,N_1,N_2,f^*)$, irrespective of the number of boundary conditions.  Each parameter of state defines one independent  way of changing the 
internal energy of the system. The parameter $f^*$ is responsible for additional net heat, not included in $T^*dS^*$. 
$T^*dS^*$ is not the differential form describing total net heat  in the system. The total net heat is given not only by the changes in $S^*$ but also in $f^*$.
The same observation was made for the van der Waals gas in the heat flow. In this case the total net heat was given not only by $S^*$, but also by the renormalized in the mapping procedure of two parameters of state $a^*,b^*$ describing attractive interactions and excluded volume in the van der Waals gas~\cite{Karol}.  
These parameters are constant at equilibrium since they are material parameters that define interactions in a particular system, but in non-equilibrium van der Waals gas they are parameters of state, which change the energy due to the change of density profiles. In our case of binary mixture the new parameter of state emerged from concentration profiles. 
Due to existence of  profiles of different physical quantities, we expect  all material parameters  to become state parameters  in the non-equilibrium systems.

\textbf{Author Contributions}

Conceptualization, R.H., A.M. and K.M.; methodology, R.H., K.M., A.M. and P.J.{\.Z}.; formal analysis, R.H., A.M. and K.M.; investigation, R.H., A.M.,  K.M., K.G. and P.J.{\.Z}; writing-original draft preparation, A.M.; writing-review and editing, R.H., K.M., A.M., K.G. and P.J.{\.Z}.. All authors have read and agreed to the published version of the manuscript.

\textbf{Acknowlegements}

A.M. was supported by the Polish National Science Center (Opus Grant   No. 2022/45/B/ST3/00936).
P.J.Z. would like to acknowledge the support of a project that has received funding from the European Union's Horizon 2020 research and innovation program under the Marie Sk{\l}odowska-Curie Grant Agreement No. 847413 and was a part of an international cofinanced project founded from the
program of the Minister of Science and Higher Education entitled ``PMW'' in the years 2020-2024; Agreement No. 5005/H2020-MSCA-COFUND/2019/2.

\appendix
\section{Variables of steady state in terms of  control parameters }
\label{app:A}
The integral of the concentration profile $x_1(z)$ in   Eq.~(\ref{eq:f1}) for the integrand given by Eq.~(\ref{eq:xT}) with $N_1=N_2$  can be obtained analytically in terms of the 
 Gaussian hypergeometric function $_2F_1(a,b;c;w)$ and polygamma function $\psi^{(m)}(w)$~\cite{Abramowitz}:
\begin{equation}
 \begin{aligned}\label{app:eq1}
& \frac{1}{L}\int_0^L x_1(z)dz =\frac{1}{s\delta} \left[\psi^{(0)}(d) -\psi^{(0)}(d+1/2)\right] \nonumber \\
&+ \frac{2}{s(2+\delta)}\left(1+s\right)^{1+\frac{\delta}{2}}\; _2F_1(1,2d;2d+1;-(1+s)^{\frac{\delta}{2}}),
\end{aligned}
 \end{equation}
where $d = 1/\delta+1/2$ with  $\delta=f_1-f_2>0$  and  $s = (T_2-T_1)/T_1  >0$  is the reduced slope  of the temperature profile.
The state variable $f_1^*$ is thus
\begin{equation}
\label{app:eq2}
f_1^*=\frac{f_1}{x_1} \left[\mathcal{B}_1(\delta,s)+\mathcal{B}_2(\delta,s)\right],
\end{equation}
where
\begin{equation}
 \begin{aligned}\label{app:eq3}
\mathcal{B}_1(\delta,s) &=\frac{1}{s\delta} \left[\psi^{(0)}(d) -\psi^{(0)}(d+1/2)\right] \nonumber \\
\mathcal{B}_2(\delta,s)&= \frac{2}{s(2+\delta)}\left(1+s\right)^{1+\frac{\delta}{2}}\; _2F_1(1,2d;2d+1;-(1+s)^{\frac{\delta}{2}}).
\end{aligned}
 \end{equation}

For $N_1\ne N_2$ we obtain
\begin{equation}
\label{app:eq2}
f_1^*=\frac{f_1}{x_1} \left[\mathcal{D}_1\left(\delta,s,\frac{N_1}{N_2}\right)+\mathcal{D}_2\left(\delta,s,\frac{N_1}{N_2}\right)\right],
\end{equation}
where
\begin{equation}
 \begin{aligned}\label{app:eq3}
\mathcal{D}_1\left(\delta,s,\frac{N_1}{N_2}\right) &= - \frac{2N_1}{N_2s(2+\delta)}\; _2F_1\left(1,2d;2d+1;-\frac{N_1}{N_2}\right)\nonumber \\
\mathcal{D}_2\left(\delta,s,\frac{N_1}{N_2}\right)&= \frac{2N_1\left(1+s\right)^{1+\frac{\delta}{2}}}{N_2s(2+\delta)}\; _2F_1\left(1,2d;2d+1;-\frac{N_1}{N_2}(1+s)^{\frac{\delta}{2}}\right).
\end{aligned}
 \end{equation}
For small reduced temperature gradients 
$s = (T_2-T_1)/T_1$ across the system  we find for $N_1\ne N_2$
\begin{equation}
\begin{aligned}
\label{app:eq4}
f_1^*&\approx \frac{f_1}{x_2}\; _2F_1\left(1,2d;2d+1;-\frac{x_1}{x_2}\right) \\
&- \frac{f_1}{x_1}\left(-\frac{x_1}{x_2}\right)^{-\frac{2}{\delta}}\mathrm{B}\left(-\frac{x_1}{x_2};2d+1;-1\right) \nonumber \\
&+\frac{f_1}{x_2}\frac{f_1-f_2}{4(1+x_1/x_2)^2}s +O(s^2),
\end{aligned}
\end{equation}
where $\mathrm{B}(s;p,q)$ is incomplete Euler beta function~\cite{Abramowitz}.
For $N_1=N_2$ the above expression reduces to Eq.~(\ref{eq:f1a}).

\end{document}